\documentclass[conference]{IEEEtran}
\IEEEoverridecommandlockouts

\usepackage{cite}
\usepackage{amsmath,amssymb,amsfonts}
\usepackage{algorithmic}
\usepackage{graphicx}
\usepackage{textcomp}
\usepackage{hyperref}
\usepackage{subfig}
\usepackage{xcolor}
\def\BibTeX{{\rm B\kern-.05em{\sc i\kern-.025em b}\kern-.08em
    T\kern-.1667em\lower.7ex\hbox{E}\kern-.125emX}}
\begin{document}

\title{REF-VC: Robust, Expressive and Fast Zero-Shot Voice Conversion with Diffusion Transformers}
\author{
 \textbf{Yuepeng Jiang$^{1}$},
 \textbf{Ziqian Ning$^{1}$},
 \textbf{Shuai Wang$^{2}$},
 \textbf{Chengjia Wang$^{3}$},
 \textbf{Mengxiao Bi$^{3}$},
\\
 \textbf{Pengcheng Zhu$^{4 \ast}$},
 \textbf{Zhonghua Fu$^{1}$},
 \textbf{Lei Xie$^{1}$}\thanks{* Corresponding authors.}
\\
 $^{1}$Audio, Speech and Language Processing Group (ASLP@NPU), \\ School of Software, Northwestern Polytechnical University, Xi’an, China
 \\
 $^{2}$School of Intelligence Science and Technology, Nanjing University, Suzhou, China \\ 
 $^{3}$Fuxi AI Lab, NetEase, China \\
 $^{4}$Geely, China
\\
 \small{
   \href{jiangyp@mail.nwpu.edu.cn}{jiangyp@mail.nwpu.edu.cn}, \href{Pengcheng.Zhu6@geely.com}{Pengcheng.Zhu6@geely.com}
 }
}



\maketitle

\begin{abstract}
In real-world voice conversion applications, environmental noise in source speech and user demands for expressive output pose critical challenges. Traditional ASR-based methods ensure noise robustness but suppress prosody richness, while SSL-based models improve expressiveness but suffer from timbre leakage and noise sensitivity. This paper proposes REF-VC, a noise-robust expressive voice conversion system. Key innovations include: (1) A random erasing strategy to mitigate the information redundancy inherent in SSL features, enhancing noise robustness and expressiveness; (2) Implicit alignment inspired by E2TTS to suppress non-essential feature reconstruction; (3) Integration of Shortcut Models to accelerate flow matching inference, significantly reducing to 4 steps. Experimental results demonstrate that REF-VC outperforms baselines such as Seed-VC in zero-shot scenarios on the noisy set, while also performing comparably to Seed-VC on the clean set. In addition, REF-VC can be compatible with singing voice conversion within one model. The samples can be found at: \href{https://rxy-j.github.io/asru2025/}{https://rxy-j.github.io/asru2025/}
\end{abstract}

\begin{IEEEkeywords}
voice conversion, noise-robust, expressive, implicit alignment, flow matching, shortcut models
\end{IEEEkeywords}

\section{Introduction}

Voice conversion (VC) is a technique that transforms a speaker's voice into that of another speaker without altering the linguistic content.
VC has been widely used in various domains, including movie and game dubbing, voice chat, and other scenarios.
However, in real-world applications, noise is unavoidable in user recordings. It's crucial to ignore noise in the source speech and generate clean, high-quality human voices as output.
Meanwhile, advancements in technology have led to increasing user expectations for VC applications. Beyond preserving linguistic information, there is a growing demand to retain paralinguistic information, enabling more natural and spontaneous speech.
This requires VC systems to capture and reproduce aspects such as tone, emotion, and even non-verbal elements like laughter and crying, ensuring more realistic and expressive output.

To address the first issue, i.e., noise robustness, previous studies have explored the use of adversarial learning~\cite{noiserobustvcat, noiserobustvcdat} or data augmentation~\cite{noro, tdrvc} to disentangle noise from the input speech. 
However, these approaches face significant limitations when dealing with unseen types of noise.
Another common solution leverages a well-trained automatic speech recognition (ASR) model, which inherently exhibits a certain degree of noise robustness due to its training objective~\cite{overviewnrasr, asrsurvey}.
From early phonetic posteriorgram (PPG)-based approaches~\cite{ppgvc, m2mppgvc} to more recent bottleneck feature (BNF)-based methods~\cite{accentvc, dualvc}, ASR-based content modeling has consistently demonstrated stable performance in VC tasks.
Nevertheless, the main drawback of these models is that the ASR training objective overly emphasizes linguistic content while heavily suppressing paralinguistic information.
Although this avoids source timbre leakage and provides strong noise robustness, it also eliminates prosodic information, leading to flatter rhythms and reduced naturalness in the converted speech.

To enhance the preservation of expressiveness, researchers have adopted self-supervised learning (SSL) models, such as Wav2Vec~\cite{wav2vec2} and WavLM~\cite{wavlm}, to replace automatic speech recognition (ASR) models in VC systems~\cite{s2vc, selfvc, freevc}.
The features extracted from these models are compressed representations of audio that retain rich linguistic and paralinguistic information, thereby improving the naturalness and expressiveness of converted speech.
However, this approach introduces new challenges, such as source timbre leakage and reduced noise robustness.
To address these issues, some methods employ k-means clustering~\cite{vectokvc, skqvc} or vector quantization~\cite{takinvc} to create information bottlenecks that filter out unwanted elements.
These approaches, however, are highly sensitive to parameter settings. Improper configurations can easily result in instability in content representation or prosody.

Overall, the ASR-based model performs well in content modeling and exhibits excellent noise robustness, while SSL-based models are superior in capturing paralinguistic content.
To overcome the trade-off challenge between noise robustness and expressiveness preservation in existing voice conversion systems, we propose REF-VC-a \textbf{R}obust, \textbf{E}xpressive and \textbf{F}ast voice conversion system.
Our model adopts the diffusion transformers (DiT)~\cite{DiT} as its backbone and effectively integrates the complementary advantages of ASR and SSL models.


\begin{figure}[!tbp]
    \centering
    \includegraphics[width=\columnwidth]{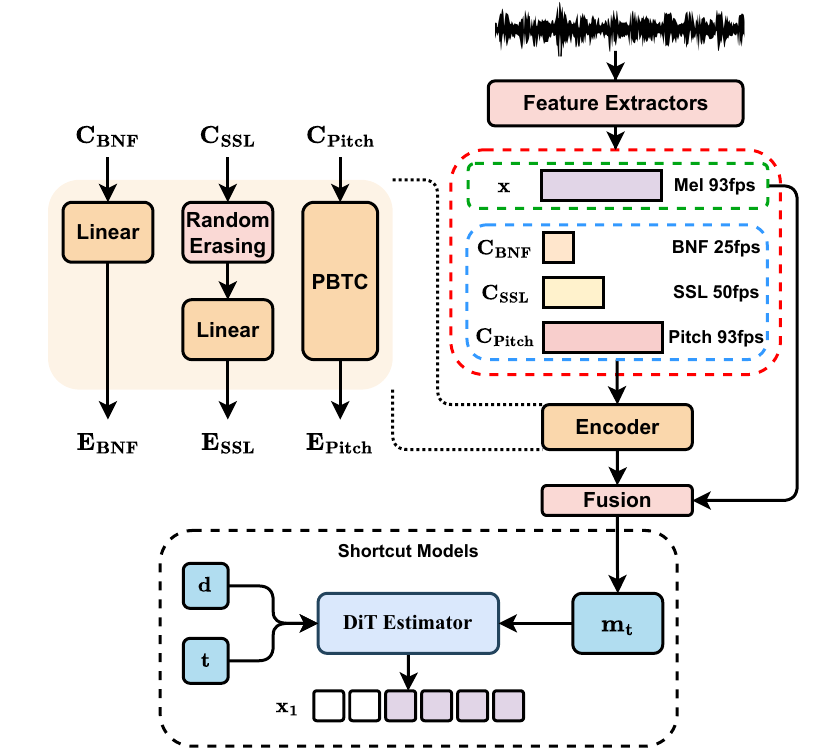}
    \caption{Architecture overview of REF-VC}
    \label{fig:system_overview}
\end{figure}

The contributions of this paper are summarized as follows,
\begin{itemize}
    \item We propose a VC system that integrates ASR and SSL features. To address the instability issues caused by redundant information in SSL features, we introduce a simple yet effective random erasing strategy. Unlike existing feature fusion approaches, our method requires neither adding perturbations to inputs nor employing information bottlenecks to resolve timbre leakage issues. This approach avoids complex model tuning and potential information loss while enhancing system noise robustness and achieving expressive voice conversion.
    \item Unlike conventional frame-to-frame conversion methods, this system employs an implicit alignment approach inspired by E2TTS~\cite{e2tts}. This alignment strategy serves to further minimize the model's reconstruction of unimportant information in the input features, thereby enhancing the quality of the conversion results.
    \item To reduce the number of inference steps of flow matching~\cite{flowmatching}, Shortcut Models~\cite{shortcut-model} is introduced. It creates shortcuts by building self-consistency properties upon flow matching.
    \item Experiments demonstrate the superiority of our proposed system. Compared to baseline models such as Seed-VC~\cite{seedvc}, our method achieves higher speaker similarity and lower character error rate on both clean and noisy sets in zero-shot voice conversion. The introduction of the Shortcut Models enables inference to be completed in just 4 steps. Additionally, REF-VC is compatible with singing voice conversion\footnote{Samples can be found on demo page} within a single model.
    \item The pretrained models trained on large-scale datasets, and the complete training recipe will be publicly available.
\end{itemize}

\begin{figure}[!tbp]
    \centering
    \includegraphics[width=\columnwidth]{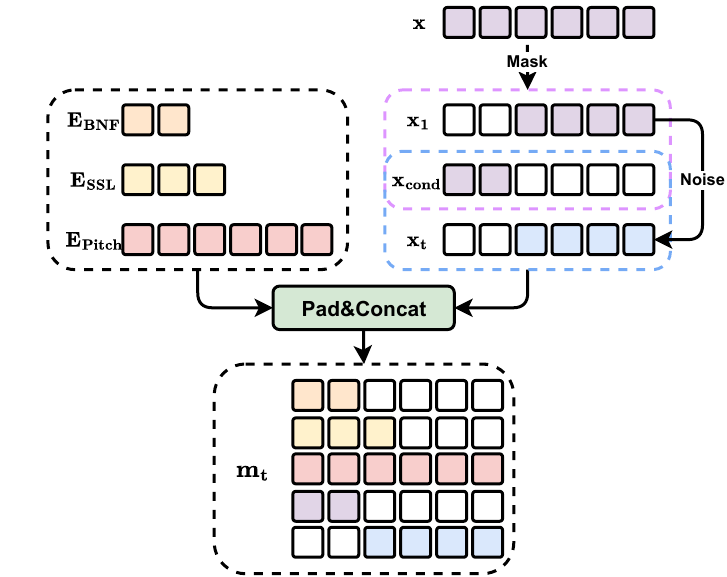}
    \caption{Detail of fusion module.}
    \label{fig:fusion}
\end{figure}

\section{Proposed Approach}

\subsection{Overview}


As illustrated in Figure~\ref{fig:system_overview}, our proposed model comprises three core components: an input encoder, fusion module, and DiT-based estimator. We employ pretrained Wenet~\cite{wenet} to extract bottleneck features $C_{BNF}$ and utilize WAVLM~\cite{wavlm} to extract self-supervised representations $C_{SSL}$. The input encoder projects these $C_{BNF}$ and $C_{SSL}$ features into a low-dimensional latent space as content conditioning. For fundamental frequency $C_{pitch}$ extracted from audio, we implement multi-scale pitch modeling through Parallel Biased Transposed Convolution (PBTC) modules~\cite{pbtcsvc, pbtcsvcc2023}. Drawing inspiration from E2TTS frameworks, we specifically design a VC-optimized fusion module to generate the estimator input $m_t$. The estimator is trained based on flow matching. Furthermore, we incorporate Shortcut Models to accelerate the sampling process while maintaining high-quality synthesis.


\subsection{Random Erasing Strategy}

Constrained by ASR training objectives, BNF contains rich linguistic but lacks paralinguistic information.  As a compressed representation of audio, SSL features compensate for this shortcoming in BNF. Incorporating SSL features alongside BNF as model inputs can effectively enhance the paralinguistic content performance (e.g., prosody) in converted speech. However, the rich information within SSL features leads the model to overly rely on them for audio modeling, which is detrimental not only to timbre similarity but also to noise robustness. Ideally, with the introduction of SSL features, the model should be able to rely primarily on BNF for audio modeling while only utilizing useful information from the SSL features.

We propose a simple yet effective random erasing strategy to regulate feature attention allocation. During training, we randomly replace SSL features with noise through batch-wise erasure operations. For each individual sample in a batch, the erasure probability ranges from 0 to 1, while maintaining an overall batch erasure probability of 0.5. This mechanism effectively suppresses the model's reliance on SSL features, compelling it to primarily utilize BNF for audio reconstruction, thereby preserving robust noise-resistant capabilities. For SSL features, the model focuses on the content in them that contributes to model convergence. In our task, this refers to paralinguistic content. Therefore, the random erasure strategy does not degrade the paralinguistic performance (e.g., prosody and emotional expression) of generated audio. Meanwhile, with the support of the random erasure strategy, the timbre leakage issue induced by SSL features has also been significantly mitigated.

\subsection{Implicit Alignment for Voice Conversion}



The frame-level input-output alignment characteristics in ASR-based or SSL-based voice conversion models merit particular attention. Conventional approaches typically employ transposed convolutions or interpolation to reconcile the frame rate mismatch between input content features and training targets like mel-spectrograms. This alignment strategy significantly reduces the model's difficulty in reconstructing speech-irrelevant content within current frames, thereby potentially compromising audio clarity and noise robustness. In contrast, text-to-speech (TTS) systems conventionally address alignment challenges by mapping unaligned content to averaged representations (typically silence). Implementing similar alignment mechanisms in voice conversion could mitigate noise robustness degradation caused by over-reconstruction. Methods like StableVC~\cite{stablevc} introduce alignment through input feature quantization and repetitive token elimination, yet confront a critical trade-off between token repetition rate and codebook dimensions. Oversized codebooks yield insufficient token repetition rates that nullify alignment effectiveness, while undersized codebooks achieve higher repetition rates at the expense of potential content information loss. Furthermore, feature quantization inherently incurs unavoidable paralinguistic information degradation.

As shown in Figure~\ref{fig:fusion}, our framework introduces an implicit alignment mechanism for voice conversion via a feature fusion module inspired by E2TTS. We employ blank frame padding to extend the length of the encoder outputs $E_{BNF}$ and $E_{SSL}$ h to match the length of $x_t$. Since $E_{Pitch}$ has the same frame rate as $x_t$, it has the same length and does not need to be padded. These processed features are concatenated along the channel dimension with $x_{cond}$ and $x_t$, forming the composite fusion feature $m_t$. The estimator subsequently generates the target sequence $x_1$ conditioned on $m_t$, timestep $t$, and step size $d$ through iterative denoising.

\subsection{Shortcut Models}

In practice, in addition to the performance of the model, the speed of inference is also a key concern.
Diffusion models often require dozens of sampling steps to achieve high quality. 
This greatly increases the inference complexity of the model. 
In this paper, we choose to use Shortcut Models to speed up our model.

\begin{figure}[!t]
    \centering
    \includegraphics[scale=0.6]{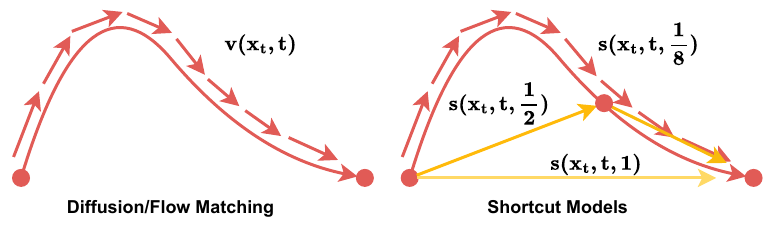}
    \caption{Comparison of shortcut models and flow matching.}
    \label{fig:shortcutmodel}
\end{figure}

Flow matching learns a path from noise to data based on ODE. 
This path is often curved. 
Sampling directly with fewer steps would lead to larger errors. 
As shown in Figure~\ref{fig:shortcutmodel}, shortcut models introduce step size $d$ to flow matching, which allows the model to adjust the direction of momentum according to $d$. This allows the model to jump to the next point as much as possible instead of deviating from the path. 
Shortcut Models is equivalent to flow matching when $d \to 0$. 
For the Shortcut Models $s_{\theta}(x_t, t, d)$, the sampling process is
\begin{equation}
x_{t+d} = x_t + s_{\theta}(x_t, t, d)d.
\end{equation}
This definition allows us to derive the inherent self-consistency of the Shortcut Models. Once this property is derived, we can transition the model from multi-step sampling to fewer steps and then to one-step sampling.
\begin{equation}
s(x_t, t, 2d) = s(x_t, t, d) / 2 + s(x_{t+d}, t + d, d) / 2.
\end{equation}
The complete loss of the Shortcut Models is as follows:
\begin{equation}
\begin{gathered}
\mathcal{L} = E[
\underbrace{||s_{\theta}(x_t, t, 0)-(x_1-x_0)||^2}_{\text{Flow-Matching}}]
\\ +E[\underbrace{||s_{\theta}(x_t, t, 2 d)-s_{\text{target}}||^2}_{\text{Self-Consistency}}], \\
\text{where} \quad s_{\text{target}}=s_{\theta}(x_t, t, d) / 2+s_{\theta}(x_{t+d}^{\prime}, t, d) / 2 \quad \\ \text{and} \quad x_{t+d}^{\prime}=x_t+s_{\theta}(x_t, t, d)d.
\end{gathered}
\end{equation}
For our model $s_{\theta}(m_t, t, d)$, the sampling procedure needs to be changed to
\begin{equation}
x_{t+d} = x_t + s_{\theta}(m_t, t, d)d, 
\end{equation}
where $m_t$ is as follows
\begin{equation}
m_t = Concat(E_{BNF}, E_{SSL}, E_{Pitch}, x_{cond}, x_t).
\end{equation}

The loss of Shortcut Models consists of two parts: flow matching loss and self-consistency loss.
Flow matching loss determines the base path of the model, while self-consistency loss is responsible for building shortcuts. 
In our training, we split a batch to calculate the two parts of the loss. 
However, in the early stage of training, the path predicted by the model is not correct.
The assumption of the shortcut does not hold.
Therefore, we do not calculate the self-consistency loss in the early stage of training.
After the flow matching loss is reduced to a certain degree, we gradually increase the proportion of self-consistency loss in a batch until the proportion reaches 1/4.

\begin{table*}[!tbp]
\centering
\caption{Objective and subjective evaluation results of comparison and ablation systems for zero-shot voice conversion. \textbf{Bold} and \underline{Underline} values indicate the best and second best results.}
\label{tab:table1}
\resizebox{\linewidth}{!}{
\begin{tabular}{l|cccc|cccc}
\hline
 & \multicolumn{4}{c|}{Clean Set} & \multicolumn{4}{c}{Noisy Set} \\ \hline
 & NMOS ($\uparrow$) & SMOS ($\uparrow$) & CER ($\downarrow$) & SECS ($\uparrow$) & NMOS ($\uparrow$) & SMOS ($\uparrow$) & CER ($\downarrow$) & SECS ($\uparrow$) \\ \hline
VITS-VC                     & 3.52$\pm$0.04 & 3.15$\pm$0.05 & 6.42 & 0.7159 & 2.84$\pm$0.03 & 2.94$\pm$0.04 & 14.17 & 0.6621 \\
Seed-VC (32NFE)               & 3.87$\pm$0.04 & \textbf{4.03$\pm$0.03} & \textbf{5.16} & \underline{0.8226} & \textbf{3.76$\pm$0.03} & 3.75$\pm$0.04 & 12.45 & 0.7884 \\
REF-VC (32NFE)                  & \textbf{3.92$\pm$0.03} & \underline{3.98$\pm$0.05}  & \underline{5.34} & \textbf{0.8253} & \underline{3.68$\pm$0.04} & \textbf{3.84$\pm$0.05} & \textbf{8.03}  & \textbf{0.8031} \\
REF-VC (4NFE)                   & \underline{3.89$\pm$0.04} & 3.87$\pm$0.05 & 5.53 & 0.8075 & 3.63$\pm$0.04 & \underline{3.78$\pm$0.06} & \underline{8.79}  & \underline{0.7919} \\ \hline
w/o implicit alignment      & 3.86$\pm$0.06 & 3.51$\pm$0.03 & 5.07 & 0.7829 & 3.24$\pm$0.04 & 3.55$\pm$0.05 & 9.97 & 0.7573  \\
w/o random erasing strategy & 3.74$\pm$0.04 & 2.77$\pm$0.08 & 4.62 & 0.5248 & 2.78$\pm$0.04 & 2.37$\pm$0.04 & 18.64 & 0.4384 \\ \hline
\end{tabular}
}
\end{table*}

\section{Experiments}

\subsection{Dataset}

We use Emilia~\cite{emilia} as the training dataset, which contains about 100,000 hours of speech data covering a wide range of speaking styles and content. This is crucial for training a robust zero-shot voice conversion model.

We set up two test sets: a clean set and a noisy set. The clean set consists of 100 audio samples randomly selected from our internal test dataset, WenetSpeech~\cite{wenetspeech} and Emilia. The noisy test set consists of 50 pieces of audio recorded in real environments using everyday devices that contain environmental or background noise. For the target speakers, we randomly select 10 speakers from seed-tts-eval dataset\footnote{\url{https://github.com/BytedanceSpeech/seed-tts-eval}}. Note that all these test sets and the target speakers are unseen during training.

\subsection{Training}

The DiT of our model consists of 12 layers, 8 attention heads, and a feedforward network dimension of 768, yielding a total of 100 million parameters. We employ Wenet~\cite{wenet} for BNF extraction and Wavlm~\cite{wavlm} for SSL feature extraction. All models are trained for 1 million steps on 8 NVIDIA A100 GPUs. The total audio length per batch is 2560 seconds. The optimizer is AdamW with an initial learning rate of 2e-4, and we use the cosine decay strategy to adjust the learning rate. We use pretrained BigVGAN\footnote{\url{https://github.com/NVIDIA/BigVGAN}} model to transform the generated mel-spectrograms into audio waveforms.

\subsection{Baseline}

We compare REF-VC with two other systems: Seed-VC\footnote{\url{https://github.com/Plachtaa/seed-vc}} and an internally developed VITS-based\cite{vits} voice conversion model (VITS-VC). Seed-VC is one of the state-of-the-art open-source voice conversion systems. And it shares a similar architecture with our approach. For a fair comparison, we utilize the official 100M-parameter checkpoint pre-trained on the Emilia, which ensures equivalent experimental conditions regarding model capacity and training data. 

\subsection{Evaluation Metrics}


For the objective evaluation, we assess two aspects: speaker similarity and intelligibility, using speaker embedding cosine similarity (SECS) and character error rate (CER), respectively. We use resemblyzer\footnote{\url{https://github.com/resemble-ai/Resemblyzer}} to evaluate SECS. CER is evaluated using the toolkit provided by seed-tts-eval.

For the subjective evaluation, we use Mean Opinion Score (MOS) to assess two aspects: speaker similarity (SMOS) and speech naturalness (NMOS). To evaluate model performance in paralinguistic reconstruction, we conduct ABX tests on three models using the clean set. Listeners are tasked to select the sample closest to the source in prosody while retaining non-verbal elements like laughter and sighs.

\section{Results}

\subsection{Subjective Evaluation}

In terms of speaker similarity, our model shows comparable performance to Seed-VC in zero-shot scenarios, while significantly outperforming VITS-VC. Experimental results show a marginal performance gap between 4-step and 32-step sampling configurations. It is noteworthy that the results on the noisy set show slightly lower similarity scores compared to the clean set, which is mainly due to the presence of samples with ambiguous pronunciation in the noisy set.

In terms of speech naturalness, our model achieves significant improvements over Seed-VC. Taking advantage of the rich information encoded in the SSL features, the converted speech effectively mitigates the problems of robotic speech prosody and pitch distortion, thus achieving significantly improved naturalness performance.

\begin{figure}[!tbp]
    \centering
    \includegraphics[width=\columnwidth]{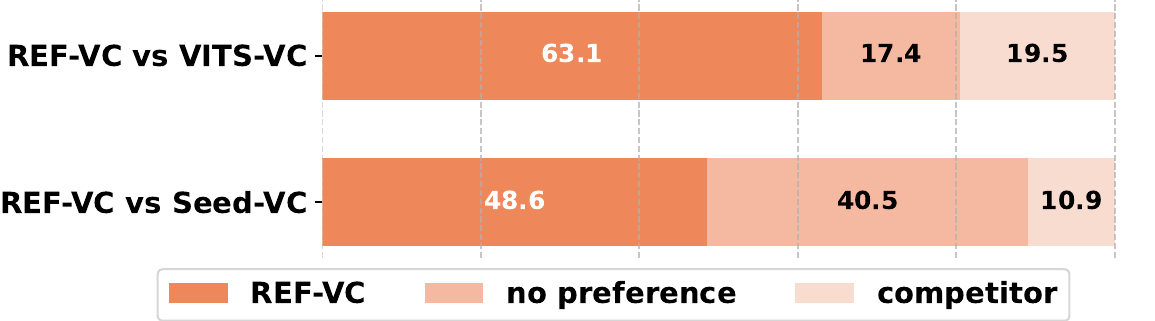}
    \caption{Result of ABX test. For Seed-VC and REF-VC, we set the number of sampling steps to 32.}
    \label{fig:abxtest}
\end{figure}

As shown in Figure~\ref{fig:abxtest}, ABX tests indicate that our model demonstrates significantly superior performance in prosody reconstruction compared to baseline models. However, it should be noted that Seed-VC's conversion results exhibit high consistency with the prosody of the prompt audio from the target speaker, and its prosodic stability decreases when the prompt audio contains distinctive stylistic characteristics.

\subsection{Objective Evaluation}


Our experimental results demonstrate significant performance differences across noise conditions. While maintaining comparable speaker similarity to Seed-VC on clean test sets, our system achieves superior performance on the noisy set. Both systems achieve an equivalent CER on the clean set; however, Seed-VC exhibits significantly degraded performance on the noisy set. Notably, despite Seed-VC being an ASR-based voice conversion benchmark, these differences validate the enhanced effectiveness of our method in preserving speech clarity in real-world scenarios involving environmental noise.

Experimental results show 4-step sampling performs slightly inferior to 32-step due to minor audio quality degradation. However, both configurations demonstrate comparable performance on subjective and objective metrics, indicating minimal speech content divergence.

\subsection{Ablation Study}

We conduct ablation studies on two key designs: the random erasing strategy and implicit alignment. For the random erasing strategy ablation, we set the random erasing ratio to 0 while maintaining implicit alignment. In the implicit alignment ablation, we implement feature alignment through interpolation.

As demonstrated in Table~\ref{tab:table1}, the removal of the random erasing strategy results in significant degradation of both audio quality and speaker similarity, confirming its dual functionality in not only reducing the model's attention to speech-irrelevant patterns in SSL features but also substantially mitigating voice timbre leakage induced by SSL features.

\begin{figure}[!tbp]
    \centering
    \subfloat[ground truth]{%
        \includegraphics[width=0.23\textwidth]{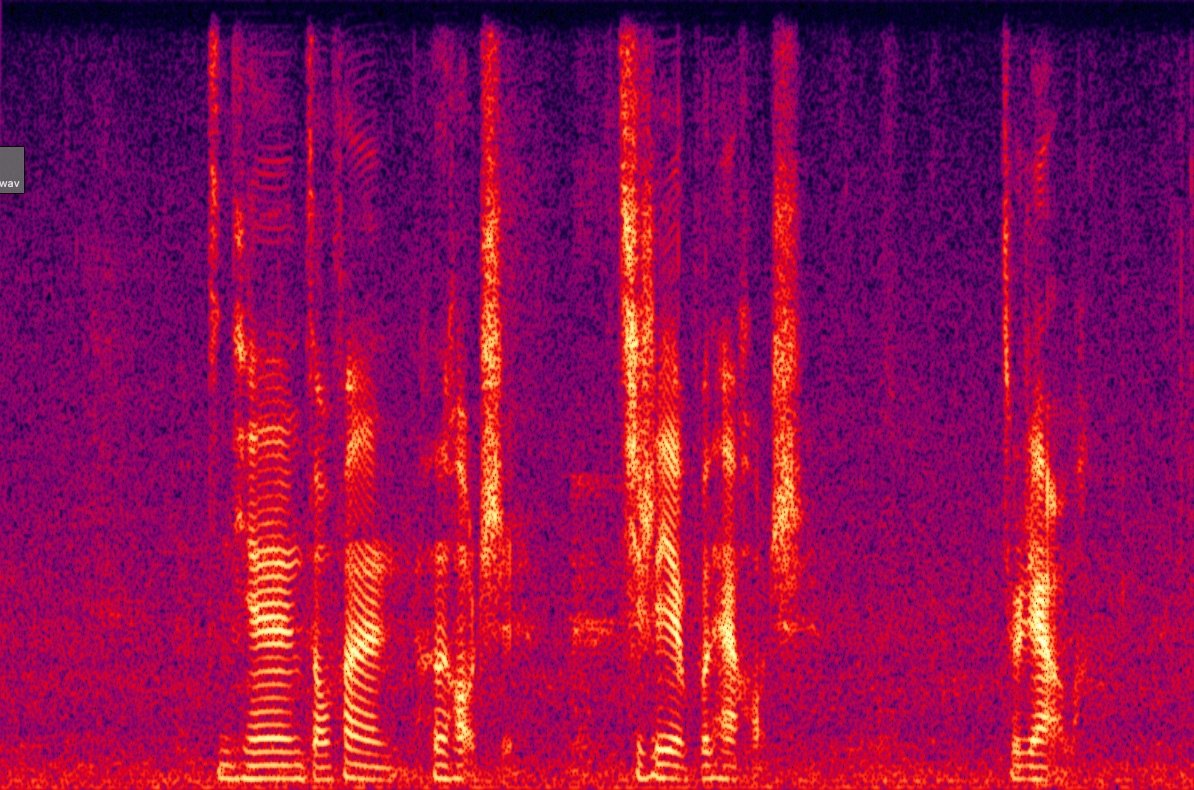}%
        \label{fig:sub-a}%
    }
    \hfill
    \subfloat[w/o random erasing strategy]{%
        \includegraphics[width=0.23\textwidth]{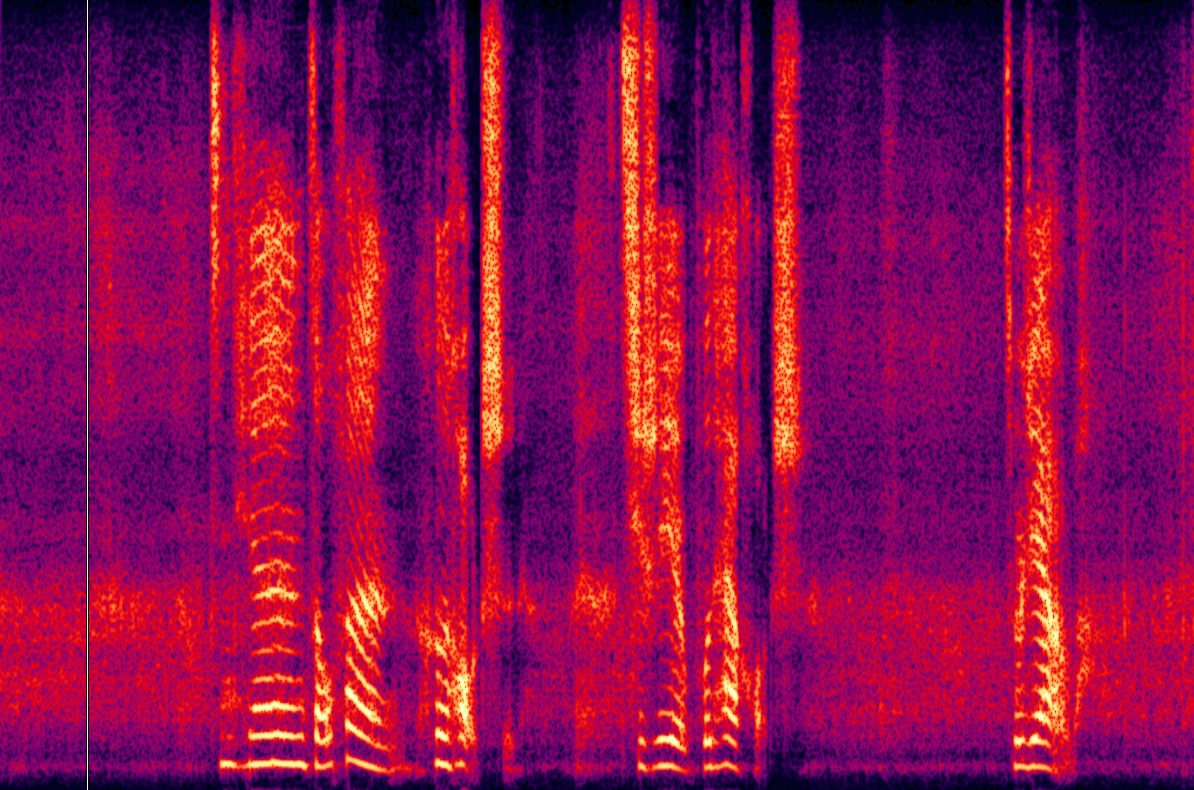}%
        \label{fig:sub-b}%
    }
    \hfill
    \subfloat[w/o implicit alignment]{%
        \includegraphics[width=0.23\textwidth]{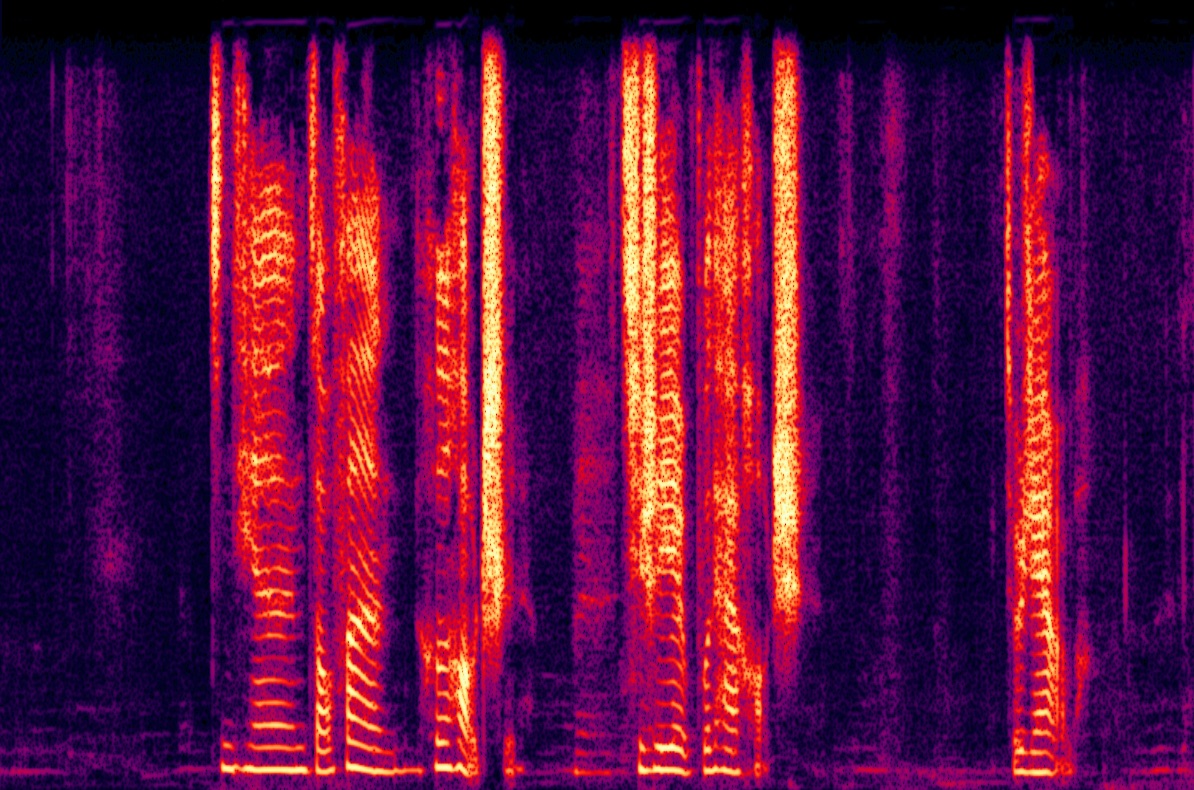}%
        \label{fig:sub-c}%
    }
    \hfill
    \subfloat[proposed model]{%
        \includegraphics[width=0.23\textwidth]{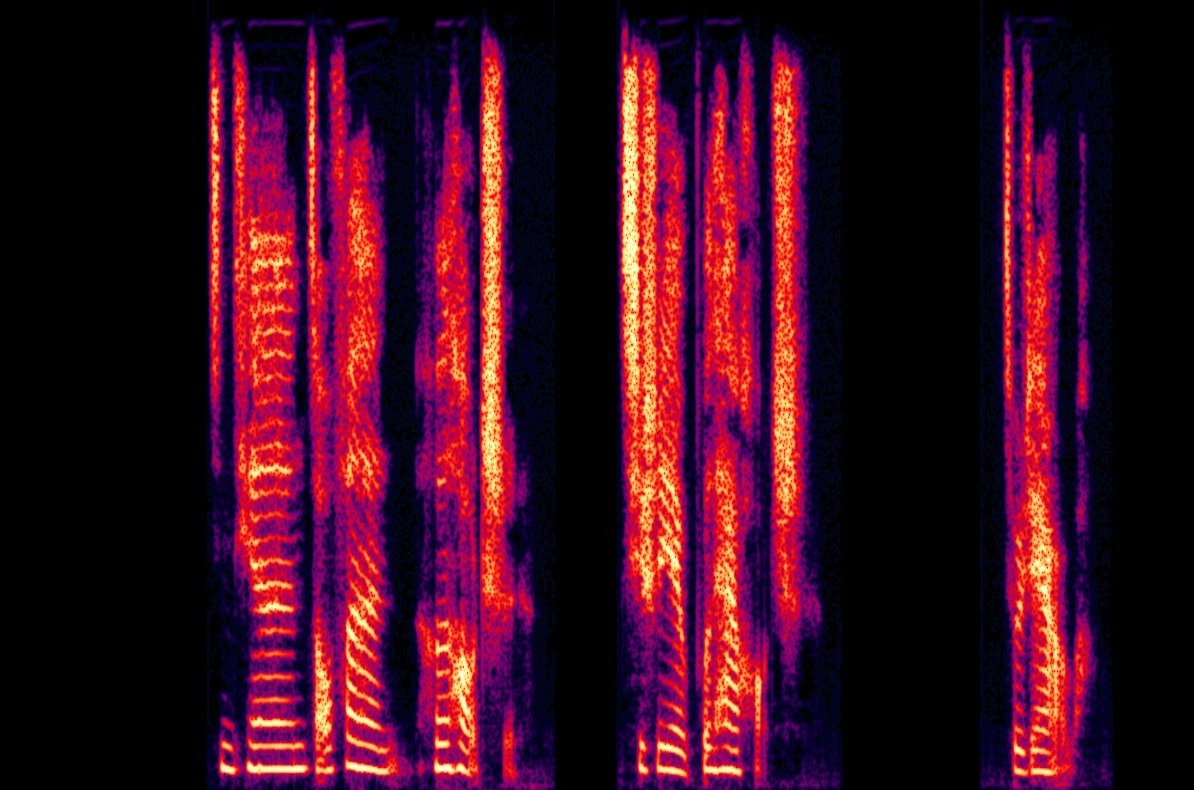}%
        \label{fig:sub-d}%
    }
    \caption{Spectrogram visualization of ablation experiments.}
    \label{fig:compare}
    \vspace{-8pt}
\end{figure}

Furthermore, experimental results reveal that incorporating implicit alignment further enhances audio reconstruction quality. The visual comparison in Figure~\ref{fig:compare} illustrates the quality improvement achieved by our proposed strategy. The ground truth audio shown in Figure~\ref{fig:compare}~(a) contains noticeable background noise. In the ablation study of the random erasing strategy (Figure~\ref{fig:compare}~(b)), the generated audio exhibit noticeable background noise. The ablation study of implicit alignment (Figure~\ref{fig:compare}~(c)) still demonstrates residual background noise reconstruction. In contrast, the audio reconstructed by our proposed model (Figure~\ref{fig:compare}~(d)) achieves  almost complete elimination of background noise.


\section{Conclusions}


This paper proposes REF-VC, a noise-robust zero-shot voice conversion model that effectively combines BNF and SSL features via random erasing strategy to improve noise robustness while maintaining expressiveness. An implicit alignment mechanism enhances audio fidelity in challenging environments. Experiments show comparable performance to state-of-the-art open-source models on clean set and superior results on noisy set. Notably, by introducing Shortcut Models, we reduce sampling steps from 32 to 4 with minimal quality loss.

\section{Future Works}


In our experiments, we observe that Seed-VC demonstrates capability in transferring target speaker styles, which contributes to improved speaker similarity. In contrast, our proposed model prioritizes faithful preservation of source prosody. Our design not only ensures good naturalness but also inherently supports singing voice conversion. However, through practical investigations, we identify that users generally prefer models capable of simultaneously converting both timbre and style. Future work will focus on investigating approaches to concurrently achieve prosody preservation and style transfer.

Moreover, unlike conventional VC systems, our model cannot generate arbitrarily long speech. Our model performs similarly to TTS systems~\cite{e2tts, f5tts}. The introduction of implicit alignment prevents our model from synthesizing excessively long audio. We will address this duration limitation in future research.

\bibliographystyle{IEEEtran}
\bibliography{mybib}

\end{document}